\documentclass[referee]{aa}

\usepackage{graphicx}
\usepackage{txfonts}
\usepackage{lipsum}
\usepackage{subcaption}         
\usepackage{lscape}             
\usepackage{placeins}           
\usepackage{cuted}                               
\usepackage[final]{hyperref}

\begin{document}


   \title{2026 QQ$_8$: Earth's temporary flyby of maximum lunation}

%
%
%



   \author{
   Grigori Fedorets\inst{1,2}\corrauth{fedorets@iki.fi}
   \and Marco Micheli\inst{3}\email{marco.bs.it@gmail.com}
   \and Samantha Hemmelgarn\inst{4}\email{shemmelgarn@lowell.edu}
   \and Nicholas Moskovitz\inst{4}\email{nmosko@lowell.edu}
   \and Krisztián Sárneczky\inst{5}\email{sarneczky.krisztian@csfk.org}
   \and David J. Tholen\inst{6}\email{tholen@hawaii.edu}
   \and Anlaug Amanda Djupvik\inst{7}\email{amanda@not.iac.es}
   \and Mikael Granvik\inst{2,8}\email{mgranvik@iki.fi}
   \and Bill Gray\inst{9}\email{pluto@projectpluto.com}
   \and Robert Jedicke\inst{6}\email{jedicke@hawaii.edu}
   \and William H. Ryan\inst{10}\email{Bill.Ryan@nmt.edu}
   \and Eileen V. Ryan\inst{10}\email{eileen.ryan@nmt.edu}
   } 

   \institute{
    Finnish Centre for Astronomy with ESO, University of Turku, Vesilinnantie 5, 20014 Turun yliopisto, Finland
    \and Department of Physics, University of Helsinki, P. O. Box 64, 00014 Helsingin yliopisto, Finland
    \and ESA NEO Coordination Centre, Planetary Defence Office, European Space Agency, Largo Galileo Galilei 1, 00044, Frascati, RM, Italy
    \and Lowell Observatory, 1400 W Mars Hill Rd, 86001, Flagstaff, AZ, USA
    \and Konkoly Observatory, HUN-REN Research Centre for Astronomy and Earth Sciences, H-1121 Budapest, Konkoly Th.M. 15-17, Hungary
    \and Institute for Astronomy, University of Hawai'i, 2680 Woodlawn Drive, 96822, Honolulu, HI, USA
    \and Nordic Optical Telescope, Rambla José Ana Fernández Pérez 7, 38711, Breña Baja, Spain
    \and Asteroid Engineering Laboratory, Luleå University of Technology, Bengt Hultqvists väg 1, 98192, Kiruna, Sweden
    \and Project Pluto, 168 Ridge Rd, 04008 Bowdoinham, ME, USA
    \and Magdalena Ridge Observatory, New Mexico Tech, 801 Leroy Pl, 87801, Socorro, NM, USA
   }

   \date{Received November xx, 20XX}

 
  \abstract
   {}
   {On a rare occasion, asteroids are temporarily captured by the Earth to become the planet's interim companions. The discovery of asteroid 2026~QQ$_8$ provided an opportunity to study this class of objects, adding to the sparse discovery catalogue of Earth's temporary companions.} 
   {We performed astrometric, photometric, and dynamical studies of 2026~QQ$_8$, and measured its solar radiation pressure signal to discern between a natural and an artificial origin.}
   {We find 2026 QQ$_8$ to be of natural origin, with solar radiation pressure signal commensurable with natural objects and spectrophotometry of an S-type object. 2026 QQ$_8$ is Earth's third discovered otemporarily-captured companion on a flyby orbit, and the first long-duration one, owing this to being on the so-called "orbit of maximum lunation". Its unusual orbit lies close to the theoretical upper duration limit of an object captured in the Earth-Moon system, and has now been observed for the first time. }
   {}

   \keywords{Minor planets, asteroids: individual: 2026 QQ$_8$; Astrometry and celestial mechanics;}

   \maketitle
   \nolinenumbers

\section{Introduction} \label{s:intro}

Asteroids occasionally become temporarily captured by planets, including the Earth. Temporary companions are divided into two dynamical classes of orbiters and flybys on the basis of whether they complete a full orbit during the capture event. Until now, the discoveries include two temporary orbiters, 2006~RH$_{120}$ \citep{kwiatkowski2009}, and 2020~CD$_3$ \citep{fedorets2020b,bolin2020,naidu2021}; and two temporary flybys,  2022~NX$_1$ \citep{delafuentemarcos2023} and 2024~PT$_5$ \citep{delafuentemarcos2024,delafuentemarcos2025,bolin2025,kareta2025}. Also, two recorded meteors have had geocentric orbits \citep{clark2016,shober2019}. The observed discovery rate of temporary companions, is currently in agreement with theoretical models \citep{granvik2012,fedorets2017}, which predict that at any given time, an asteroid with a diameter of about 80\,cm is captured in a geocentric orbit while an object with a diameter of 3\,m is captured once every 10--20 years.. 

Earth's temporary companions play an important role in population studies of small near-Earth objects (NEOs). There is an order-of-magnitude mismatch between the theoretical predictions of metre-sized NEOs  \citep{granvik2018,harris2021,nesvorny2024}, and the observations of metre-sized Earth-impactors as bolides \citep{brown2013,chow2025}. Since obtaining a statistically-significant sample of metre-sized NEOs is a long and arduous task, the discovery cadence of Earth's temporary satellites is an early indicator of the relative abundance of metre-sized NEOs. The population of temporary satellites is also hypothesised to include lunar ejecta \citep{gladman1995,jedicke2025}, and can thus provide constraints for lunar cratering mechanisms. Finally, due to their close proximity to the Earth, temporary satellites are viable targets for testing in situ resource utilisation of asteroids \citep{granvik2013,jedicke2018}.

Earth's temporary satellites pose an edge case in automatic classification of objects. As the overwhelming majority of objects on geocentric orbits are artificial, they are typically immediately removed from the Minor Planet Center's NEO Confirmation Page (NEOCP). Following this main rule, the removal of Earth's natural satellites has occurred in the majority of cases where such objects have been discovered. With the expected surge in temporary natural satellite discoveries \citep{fedorets2020a} by the Vera C. Rubin Observatory's Legacy Survey of Space and Time \citep[LSST; ][]{ivezic2019}, it has become evident that a procedure to process suspected temporary satellite candidates must be established.

Here, we present the results of the observational campaign for asteroid 2026~QQ$_8$ (MPEC 2026-U95), an asteroid temporarily captured by the Earth in 2026. We present our astrometric and photometric observations in Sect.~\ref{s:obs}, discuss and interpret our results in Sect.~\ref{s:results}, and offer our conclusions in Sect.~\ref{s:conclusions}.


\section{Observations} \label{s:obs}

\subsection{Discovery and early astrometric follow-up}

Asteroid 2026~QQ$_8$, hereafter QQ8, was discovered by Krisztián Sárneczky with the 60-cm Schmidt telescope at the Piszkéstető Mountain Station of the Konkoly Observatory in Hungary \citep[][MPC observatory code K88]{szabo2026}, using synthetic tracking methods \citep{velkei2026} on 19 Aug 2026. At the time, the object received the temporary designation Sar2901. It was immediately recognised to be on a distant geocentric orbit, without being clearly attributable to any known distant artificial satellite or a space debris object. The discovery of potentially natural object rapidly triggered an astrometric follow-up campaign. Over the period of 19-23 Aug, additional astrometric observations were obtained from the discovery station, as well as the Visnjan observatory (L01), the Lowell Discovery Telescope (G37), Las Cumbres Observatory Tenerife stations A and B (Z24 and Z31), Schiapparelli main and southern observatories (204 and M21), and Wide-field Mufara telescope (M57).

\subsection{Extended astrometric follow-up}

As mentioned in Sect.~\ref{s:intro}, temporary satellites have often been removed from the NEOCP after discovery. Sar2901 was no exception; it was removed twice from the NEOCP, remaining absent from late August until the publication of MPEC 2026-U95 on October 29th 2026, when the provisional designation 2026~QQ$_8$ was bestowed upon the asteroid. This disrupted the workflow of several follow-up facilities which typically rely on services that either directly or indirectly rely on MPC for performing small-body observations. Fortunately, the astrometry was distributed by the observers directly to the community through the Minor Planet Mailing List. The direct approach allowed for follow-up observations of QQ8 in Sep--Oct by 2--4-m class telescopes, which were required at the time to observe a 23--25 magnitude object. Successful follow-up observations were performed with the 4.3-m Lowell Discovery Telescope (G37) and the University of Hawai'i 2.2-m telescope (T12). Multiple observations QQ8 were also attempted with the Liverpool Telescope (J13), Magdalena Ridge Observatory (H01), and Nordic Optical Telescope (Z23) during Sep--Oct. These measurements were unsuccessful due to either bad weather, or high sky brightness on nights close to full Moon on 26 Sep.

\subsection{Photometric observations}

Photometric observations of QQ8 were obtained by the Alhambra Faint Object Spectrograph (ALFOSC) mounted on the 2.5-m Nordic Optical Telescope (Z23). Observations were carried out on 21 Aug in Sloan Digital Sky Survey (SDSS) \textit{griz} filters with non-sidereal tracking and constantly updating velocities matching that of the asteroid. Standard fields with sidereal tracking were observed immediately before and after the science images. Field stars were identified against the \textit{Gaia} Data Release 3 catalogue \citep{gaia2016a,gaia2023} and cross-matched with SDSS photometric measurements. Ten stars from each field  were selected and their aperture photometry measured using a custom IRAF script \citep{tody1986,tody1993}. These measurements solved the zero point and the atmospheric correction for each filter. In the science image, the object was identified as the only non-trailed source, and its instrumental magnitude measured with IRAF. The instrumental magnitude was then corrected to an apparent magnitude in each filter using the zero-point and atmospheric corrections.


\section{Results and discussion} \label{s:results}

\subsection{Physical characterisation}

\begin{figure}[ht!]
   \centering
   \includegraphics[width=\hsize]{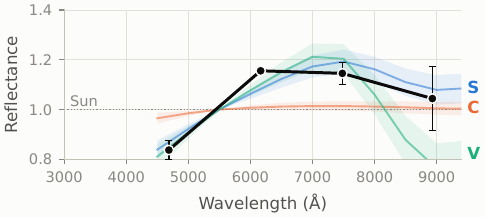}
      \caption{Normalised photometric measurements of QQ8 in the SDSS $griz$ bands as a function of wavelength, overplotted with the common types of asteroid reflectance spectra.}
         \label{fig:spectrum}
   \end{figure}

Using the solar colours by \citet{holmberg2006} and following Eq.~2 from \citet{demeo2013}, we present the measured colours of QQ8 as reflectance values in Fig.~\ref{fig:spectrum}. The taxonomic classification of the colours obtained gives the best match with the S-complex in the visual part of the Bus-DeMeo taxonomy \citep{bus2002,demeo2009}. 

Photometric modelling \citep{belskaya2000} and thermal-infrared modelling  \citep{mainzer2011} commonly yield the range of geometric albedo of atmosphereless bodies to be $p_V = 0.234 \pm 0.088$ for S-complex objects. Using Eq.~2 from \citet{pravec2007} for albedo-size relations, and $H=31.34$ from MPC, we obtain a diameter range of 1.3--1.9 m for QQ8.

\subsection{Solar radiation pressure}

\begin{figure}[ht!]
   \centering
   \includegraphics[width=\hsize]{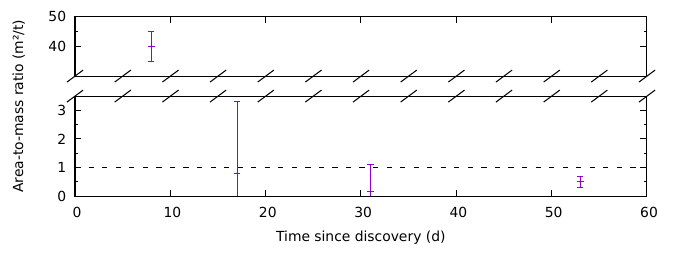}
      \caption{The evolution of the detection of the zero-albedo area-to-mass ratio with 1$\sigma$ error bars as a function of the length of the observational arc. All natural objects have had a value less than 1 m$^2$/t, indicated by the dashed line.}
         \label{fig:srp}
   \end{figure}

The evolution of the measured non-gravitational signal as a function of time is presented in Fig.~\ref{fig:srp}. We interpret the non-gravitational acceleration as solar radiation pressure (SRP), with purely radial acceleration \citep{farnocchia2015}. The latest value obtained for the area/mass ratio of QQ8 is $0.5 \pm 0.19$ m$^2$/t. The value is commensurate to other small natural objects for which solar radiation pressure values have been measured \citep{micheli2012,micheli2013,micheli2014,fedorets2020b}, and significantly below the lowest estimate for an artificial rocket booster of 6 m$^2$/t \citep{buzzoni2019}, and typical values for artificial objects \citep[e.g.,][]{früh2012}.

\subsection{Temporary capture}

\begin{figure*}[ht!]
   \centering
   \includegraphics[width=\hsize]{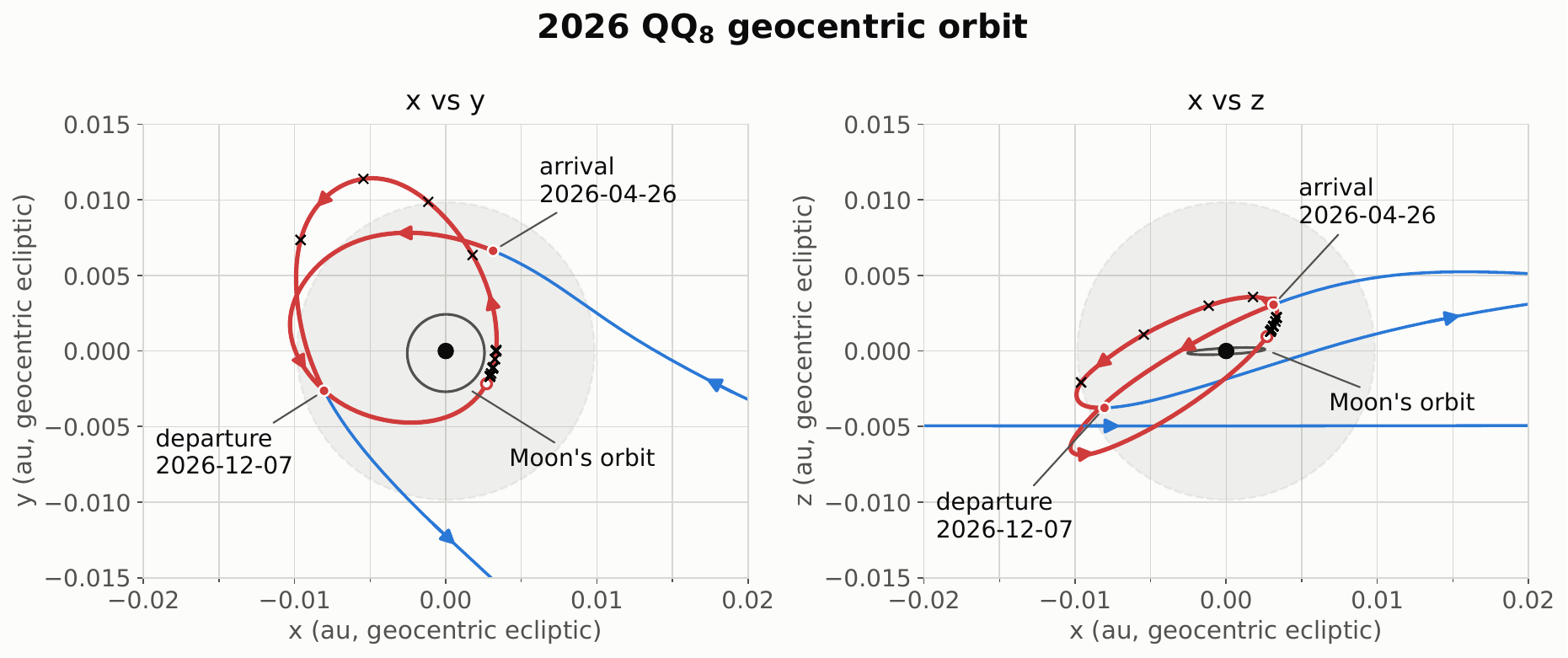}
   \includegraphics[width=\hsize]{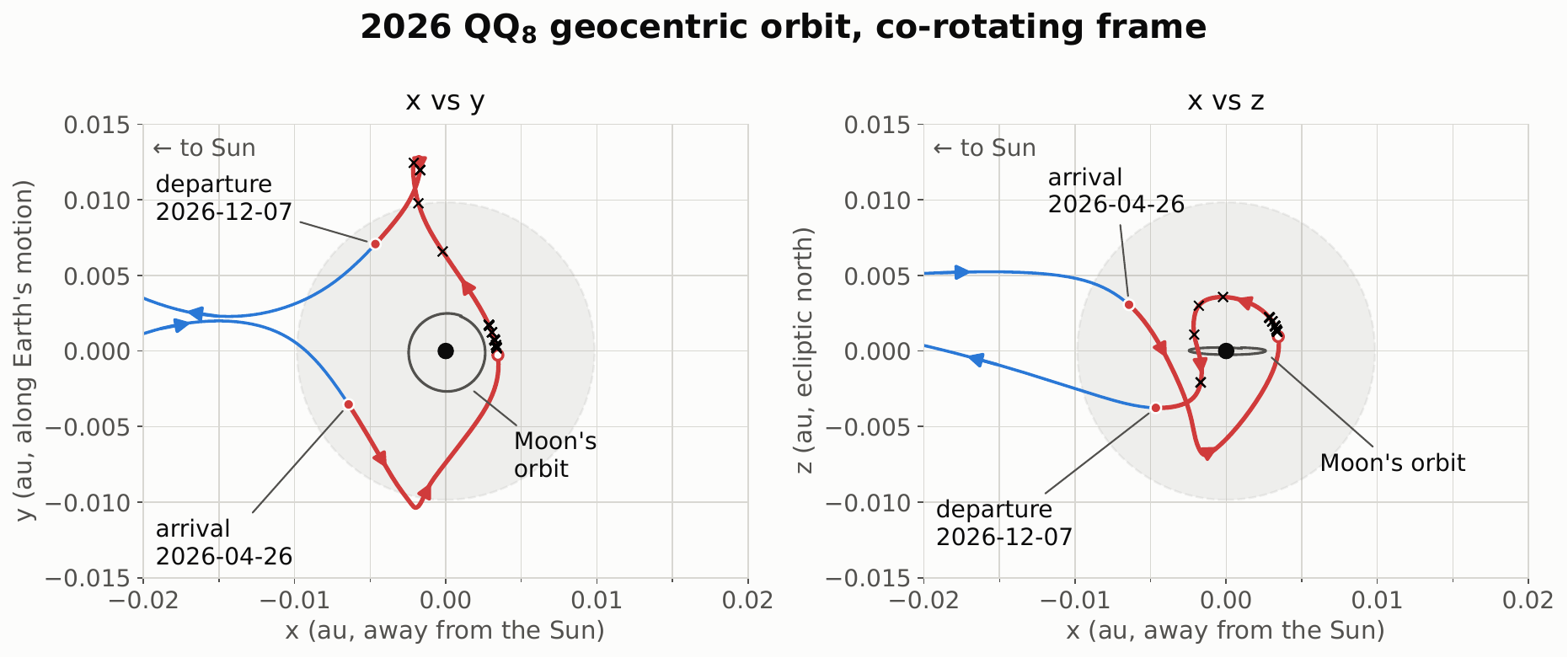}
      \caption{The geocentric capture of QQ8 in the cartesian \textit{xy} (left column) and \textit{xz} (right column) coordinates. The trajectory of QQ8 is shown in the geocentric frame (top row), and the co-rotating Sun-Earth plane (bottom row). The time during which the asteroid was captured is illustrated by the red lines, and the blue lines represent the trajectories before and after the capture. The uncertainty Monte Carlo orbital swarm is unrecognisable from underneath the main solution. The Earth's Hill sphere is shaded in grey, and the position of the Earth and the orbit of the Moon are shown. Black crosses correspond to the epochs of astrometric measurements.}
         \label{fig:orbit}
   \end{figure*}

The capture of QQ8 in the geocentric and co-rotating frame are illustrated in Fig.~\ref{fig:orbit}. Here, we fit the best solution while exploring the solution envelope until before and after the capture. For the integrations, OrbFit 5.6 \citep{orbfit2011} is used with DE440 ephemerides \citep{park2021} using the \texttt{vfcc17} astrometric weighing model \citep{veres2017}. 

QQ8 was gravitationally bound to Earth for 225 days, from 26 Apr to 7 Dec 2026. In the co-rotating coordinate system, the object fails to make a complete revolution around the Earth, which, following the definition of \citet{fedorets2017}, puts the object in the category of temporarily-captured flybys. Unlike its two predecessors, 2022~NX$_1$ and 2024~PT$_5$, that were captured for only about a month each, QQ8 has a significantly longer capture duration of 7.5~months. Moreover, the orbit of QQ8 closely resembles the "orbit of maximum lunation" postulated by \citet{hill1878}, where objects spend extended periods of time (up to 3 months) near Earth quadratures. \citet{fedorets2017} discussed these types of orbits to be responsible for the predicted overabundance of temporary satellites near Earth quadratures. The orbit of QQ8 provides the first observational evidence for the existence of such orbits.  QQ8 is among the longest 0.3\% of  flybys, 2020~CD$_3$ \citep[2.7 yr, ][]{fedorets2020b} among the longest 4\% of orbiters \citep{fedorets2017}. 
With 2 out of 5 discovered temporary companions having exceptionally long captures, the established capture mechanism requires additional scrutiny. The steady-state model remains plausible, with combined discovery interval of orbiters and flybys of 2--3 years matching  predictions. 

\subsection{Improving follow-up of Earth's natural companions}

In cases of 2006~RH$_{120}$, 2020~CD$_3$, and now QQ8, the MPC's algorithms have initially treated these objects as artificial satellites and removed them from the NEOCP as "not-asteroids" due to their geocentric orbits. The semi-automated follow-up of Earth's natural companions thus becomes disrupted, adding complications to established follow-up workflows, and making correcting for selection effects impossible unless the removal algorithm is specified. Due to typically narrow observational windows of small Earth's temporary satellites, in worst-case scenarios, we foresee such disruptions resulting in failures to obtain timely follow-up, and therefore knowledge about the true nature of these objects.

To ensure the timely follow-up of provisional natural objects on geocentric orbits, we propose several mitigation measures to be implemented in the MPC workflow upon discovery:

\begin{itemize}
    \item \textit{Applying a priori orbital information}. The geocentric Keplerian orbits of Earth's temporary satellites are typically more distant and retrograde in comparison to distant artificial satellites  / interplanetary launchers. Typical orbits of temporary satellites have been investigated by  \citet{granvik2012,fedorets2017,naidu2021}.
    \item \textit{Associating the candidate's orbit with distant artificial satellites}. As the space age is limited to time after 1957, the lack of close mutual approaches with the Earth beyond that window is a strong suggestion of the natural origin of an object on a geocentric orbit.
    \item \textit{A decreasing trend of the SRP signal}. Immediately after discovery, the error for the solar radiation pressure on the orbital fit for small objects is too large to distinguish between natural and artificial origin of a geocentric object. Both \citet{fedorets2020b} and this work have shown that for natural objects, the SRP signal is typically trending downward over a period of several weeks.
\end{itemize}

We suggest that if any of the mentioned measures indicates that the object is natural, it should remain on the NEOCP with a special flag until it is more thoroughly characterised as follows: 

\begin{itemize}
    \item \textit{Physical characterisation}. The object should be physically characterised by a combination of methods including photometry, spectroscopy, polarimetry, and/or radar. Classification should take into account both natural and artificial possibilities, e.g., for spectra of artificial objects, one can use the catalogue of \citet{battle2024}. For example, a 1 µm absorption feature is strongly indicative of natural origin.
    \item \textit{Constrained small SRP signal}. Large SRP signals indicative of artificial origin are easier to detect. For natural objects, SRP signal is typically below 1 m$^2$/t \citep[][and this work]{micheli2012,micheli2013,micheli2014,fedorets2020b}. 
\end{itemize}

The final classification decision would then to be made after going through all aforementioned steps.

\section{Conclusions} \label{s:conclusions}

\begin{itemize}
    \item We have photometrically classified 2026 QQ$_8$, an object on a geocentric orbit, to be an S-type asteroid.
    \item Astrometric measurements place the asteroid's solar radiation pressure signal to $0.5 \pm 0.19$ m$^2$/t, which is a typical value for a natural object.
    \item QQ8 is the third recorded Earth's temporary flyby, and the first flyby with a long capture of 7.5 months. A capture duration of such exceptional length is enabled by a so-called orbit of maximum lunation, where in a co-rotating frame, the object remains stationary at Earth's quadratures for extended periods of time.
    \item We propose to the MPC an algorithm enabling future facilitation of observations of natural geocentric objects.  
\end{itemize}

\begin{acknowledgements}

Based on observations made with the Nordic Optical Telescope, owned in collaboration by the University of Turku and Aarhus University, and operated jointly by Aarhus University, the University of Turku and the University of Oslo, representing Denmark, Finland and Norway, the University of Iceland and Stockholm University at the Observatorio del Roque de los Muchachos, La Palma, Spain, of the Instituto de Astrofisica de Canarias. The NOT data were obtained under program ID P70-508. The data presented here were obtained in part with ALFOSC, which is provided by the Instituto de Astrofisica de Andalucia (IAA) under a joint agreement with the University of Copenhagen and NOT. We would like to thank Astrid Lind Bouquin and Amanda Schriver Mårtensson for observational support. Data for this paper has been obtained under the International Time Programme of the CCI (International Scientific Committee of the Observatorios de Canarias of the IAC) with the Liverpool Telescope operated on the island of La Palma by the operators in the Observatorio del Roque de los Muchachos. This work has made use of data from the European Space Agency (ESA) mission
{\it Gaia} (\url{https://www.cosmos.esa.int/gaia}), processed by the {\it Gaia} Data Processing and Analysis Consortium (DPAC,
\url{https://www.cosmos.esa.int/web/gaia/dpac/consortium}). Funding for the DPAC has been provided by national institutions, in particular the institutions participating in the {\it Gaia} Multilateral Agreement. This research has made use of NASA's Astrophysics Data System Bibliographic Services. This research has made use of data and/or services provided by the International Astronomical Union's Minor Planet Center.

NM and SH acknowledge support from NASA YORPD grant 80NSSC21K1328 awarded to the Mission Accessible Near-Earth Object Survey (MANOS). These results made use of the Lowell Discovery Telescope (LDT) at Lowell Observatory. 

KS has been supported by the GINOP-2.3.2-15-2016-00003 grant of the Hungarian National Research, Development and Innovation Office (NKFIH).

MG partially supported by RCF grant \#361233.
\end{acknowledgements}

%

\bibliographystyle{aa} 
\bibliography{bibliography.bib}

@ARTICLE{brown2013,
   author = {{Brown}, P.~G. and {Assink}, J.~D. and {Astiz}, L. and {Blaauw}, R. and 
	{Boslough}, M.~B. and {Borovi{\v c}ka}, J. and {Brachet}, N. and 
	{Brown}, D. and {Campbell-Brown}, M. and {Ceranna}, L. and {Cooke}, W. and 
	{de Groot-Hedlin}, C. and {Drob}, D.~P. and {Edwards}, W. and 
	{Evers}, L.~G. and {Garces}, M. and {Gill}, J. and {Hedlin}, M. and 
	{Kingery}, A. and {Laske}, G. and {Le Pichon}, A. and {Mialle}, P. and 
	{Moser}, D.~E. and {Saffer}, A. and {Silber}, E. and {Smets}, P. and 
	{Spalding}, R.~E. and {Spurn{\'y}}, P. and {Tagliaferri}, E. and 
	{Uren}, D. and {Weryk}, R.~J. and {Whitaker}, R. and {Krzeminski}, Z.
	},
    title = {{A 500-kiloton airburst over Chelyabinsk and an enhanced hazard from small impactors}},
  journal = {Nature},
     year = {2013},
    OPTmonth = nov,
   volume = {503},
    pages = {238 -- 241},
      doi = {10.1038/nature12741},
   OPTadsurl = {http://adsabs.harvard.edu/abs/2013Natur.503..238B},
   OPTadsnote = {Provided by the SAO/NASA Astrophysics Data System}
}

@Article{clark2016,
  author = 	 {Clark, D. L. and
	            Spurn{\' y}, P. and
							Wiegert, P. and
							Brown, P. and
							Borovi{\v c}ka, J. and
							Tagliaferri, E. and
							Shrben{\' y}, L.},
  title = 	 {{Impact detection of temporarily captured natural satellites}},
  journal = 	 {Astronomical Journal},
  year = 	 {2016},
  OPTkey = 	 {},
  volume = 	 {151},
  number = 	 {6},
  pages = 	 {135-150},
  OPTmonth = 	 {},
  OPTnote = 	 {},
  OPTannote = 	 {}
  }

@Article{demeo2009,
  author = 	 {DeMeo, F. E. and
              Binzel, R. P. and
              Slivan, S. M. and
              Bus, S. J.
              },
  title = 	 {{An extension of the Bus asteroid taxonomy into the near-infrared}},
  journal = 	 {Icarus},
  year = 	 {2009},
  OPTkey = 	 {},
  volume = 	 {202},
  number = 	 {1},
  pages = 	 {160 -- 180},
  OPTmonth = 	 {},
  OPTnote = 	 {},
  OPTannote = 	 {}
  }

@Article{demeo2013,
  author = 	 {DeMeo, F. E. and
              Carry, B.},
  title = 	 {{The taxonomic distribution of asteroids from multi-filter all-sky photometric surveys}},
  journal = 	 {Icarus},
  year = 	 {2013},
  OPTkey = 	 {},
  volume = 	 {226},
  number = 	 {1},
  pages = 	 {723 -- 741},
  OPTmonth = 	 {},
  OPTnote = 	 {},
  OPTannote = 	 {}
  }

@Article{fedorets2017,
  author = 	 {Fedorets, G. and
	            Granvik, M. and
							Jedicke, R.},
  title = 	 {{Orbit and size distributions for asteroids temporarily captured by the Earth-Moon system}},
  journal = 	 {Icarus},
  year = 	 {2017},
  OPTkey = 	 {},
  volume = 	 {285},
  OPTnumber = 	 {},
  pages = 	 {83 -- 94},
  OPTmonth = 	 {},
  OPTnote = 	 {},
  OPTannote = 	 {}
  }

@Article{fedorets2020a,
  author = 	 {Fedorets, G. and
	            Granvik, M. and
	            Jones, R. L. and
	            Juri\'{c}, M. and
							Jedicke, R.},
  title = 	 {{Discovering Earth's transient moons with the Large Synoptic Survey Telescope}},
  journal = 	 {Icarus},
  year = 	 {2020a},
  OPTkey = 	 {},
  volume = 	 {338},
  OPTnumber = 	 {},
  pages = 	 {113517},
  OPTmonth = 	 {},
  OPTnote = 	 {},
  OPTannote = 	 {}
  }

@ARTICLE{fedorets2020b,
       author = {{Fedorets}, Grigori and {Micheli}, Marco and {Jedicke}, Robert and {Naidu}, Shantanu P. and {Farnocchia}, Davide and {Granvik}, Mikael and {Moskovitz}, Nicholas and {Schwamb}, Megan E. and {Weryk}, Robert and {Wierzcho{\'s}}, Kacper and {Christensen}, Eric and {Pruyne}, Theodore and {Bottke}, William F. and {Ye}, Quanzhi and {Wainscoat}, Richard and {Devog{\`e}le}, Maxime and {Buchanan}, Laura E. and {Djupvik}, Anlaug Amanda and {Faes}, Daniel M. and {F{\"o}hring}, Dora and {Roediger}, Joel and {Seccull}, Tom and {Smith}, Adam B.},
        title = "{Establishing Earth's Minimoon Population through Characterization of Asteroid 2020 CD$_{3}$}",
      journal = {\aj},
         year = {2020b},
        month = dec,
       volume = {160},
       number = {6},
          eid = {277},
        pages = {277},
          doi = {10.3847/1538-3881/abc3bc},
archivePrefix = {arXiv},
       eprint = {2011.10380},
 primaryClass = {astro-ph.EP},
       adsurl = {https://ui.adsabs.harvard.edu/abs/2020AJ....160..277F}
}

@ARTICLE{gaia2016a,
       author = {{Gaia Collaboration} and {Prusti}, T. and {de Bruijne}, J.~H.~J. and {Brown}, A.~G.~A. and {Vallenari}, A. and {624 co-authors}},
        title = "{The Gaia mission}",
      journal = {\aap},
         year = 2016,
        month = nov,
       volume = {595},
          eid = {A1},
        pages = {A1}}

@ARTICLE{gaia2023,
       author = {{Gaia Collaboration} and {Vallenari}, A. and {Brown}, A.~G.~A. and {Prusti}, T. and {de Bruijne}, J.~H.~J. and {Arenou}, F. and {Babusiaux}, C. and {Biermann}, M. and {Creevey}, O.~L. and {Ducourant}, C. and {Evans}, D.~W. and {Eyer}, L. and {Guerra}, R. and {Hutton}, A. and {Jordi}, C. and {Klioner}, S.~A. and {Lammers}, U.~L. and {Lindegren}, L. and {Luri}, X. and {Mignard}, F. and {Panem}, C. and {Pourbaix}, D. and {Randich}, S. and {Sartoretti}, P. and {Soubiran}, C. and {Tanga}, P. and {Walton}, N.~A. and {Bailer-Jones}, C.~A.~L. and {Bastian}, U. and {Drimmel}, R. and {Jansen}, F. and {Katz}, D. and {Lattanzi}, M.~G. and {van Leeuwen}, F. and {Bakker}, J. and {Cacciari}, C. and {Casta{\~n}eda}, J. and {De Angeli}, F. and {Fabricius}, C. and {Fouesneau}, M. and {Fr{\'e}mat}, Y. and {Galluccio}, L. and {Guerrier}, A. and {Heiter}, U. and {Masana}, E. and {Messineo}, R. and {Mowlavi}, N. and {Nicolas}, C. and {Nienartowicz}, K. and {Pailler}, F. and {Panuzzo}, P. and {Riclet}, F. and {Roux}, W. and {Seabroke}, G.~M. and {Sordo}, R. and {Th{\'e}venin}, F. and {Gracia-Abril}, G. and {Portell}, J. and {Teyssier}, D. and {Altmann}, M. and {Andrae}, R. and {Audard}, M. and {Bellas-Velidis}, I. and {Benson}, K. and {Berthier}, J. and {Blomme}, R. and {Burgess}, P.~W. and {Busonero}, D. and {Busso}, G. and {C{\'a}novas}, H. and {Carry}, B. and {Cellino}, A. and {Cheek}, N. and {Clementini}, G. and {Damerdji}, Y. and {Davidson}, M. and {de Teodoro}, P. and {Nu{\~n}ez Campos}, M. and {Delchambre}, L. and {Dell'Oro}, A. and {Esquej}, P. and {Fern{\'a}ndez-Hern{\'a}ndez}, J. and {Fraile}, E. and {Garabato}, D. and {Garc{\'\i}a-Lario}, P. and {Gosset}, E. and {Haigron}, R. and {Halbwachs}, J.-L. and {Hambly}, N.~C. and {Harrison}, D.~L. and {Hern{\'a}ndez}, J. and {Hestroffer}, D. and {Hodgkin}, S.~T. and {Holl}, B. and {Jan{\ss}en}, K. and {Jevardat de Fombelle}, G. and {Jordan}, S. and {Krone-Martins}, A. and {Lanzafame}, A.~C. and {L{\"o}ffler}, W. and {Marchal}, O. and {Marrese}, P.~M. and {Moitinho}, A. and {Muinonen}, K. and {Osborne}, P. and {Pancino}, E. and {Pauwels}, T. and {Recio-Blanco}, A. and {Reyl{\'e}}, C. and {Riello}, M. and {Rimoldini}, L. and {Roegiers}, T. and {Rybizki}, J. and {Sarro}, L.~M. and {Siopis}, C. and {Smith}, M. and {Sozzetti}, A. and {Utrilla}, E. and {van Leeuwen}, M. and {Abbas}, U. and {{\'A}brah{\'a}m}, P. and {Abreu Aramburu}, A. and {Aerts}, C. and {Aguado}, J.~J. and {Ajaj}, M. and {Aldea-Montero}, F. and {Altavilla}, G. and {{\'A}lvarez}, M.~A. and {Alves}, J. and {Anders}, F. and {Anderson}, R.~I. and {Anglada Varela}, E. and {Antoja}, T. and {Baines}, D. and {Baker}, S.~G. and {Balaguer-N{\'u}{\~n}ez}, L. and {Balbinot}, E. and {Balog}, Z. and {Barache}, C. and {Barbato}, D. and {Barros}, M. and {Barstow}, M.~A. and {Bartolom{\'e}}, S. and {Bassilana}, J.-L. and {Bauchet}, N. and {Becciani}, U. and {Bellazzini}, M. and {Berihuete}, A. and {Bernet}, M. and {Bertone}, S. and {Bianchi}, L. and {Binnenfeld}, A. and {Blanco-Cuaresma}, S. and {Blazere}, A. and {Boch}, T. and {Bombrun}, A. and {Bossini}, D. and {Bouquillon}, S. and {Bragaglia}, A. and {Bramante}, L. and {Breedt}, E. and {Bressan}, A. and {Brouillet}, N. and {Brugaletta}, E. and {Bucciarelli}, B. and {Burlacu}, A. and {Butkevich}, A.~G. and {Buzzi}, R. and {Caffau}, E. and {Cancelliere}, R. and {Cantat-Gaudin}, T. and {Carballo}, R. and {Carlucci}, T. and {Carnerero}, M.~I. and {Carrasco}, J.~M. and {Casamiquela}, L. and {Castellani}, M. and {Castro-Ginard}, A. and {Chaoul}, L. and {Charlot}, P. and {Chemin}, L. and {Chiaramida}, V. and {Chiavassa}, A. and {Chornay}, N. and {Comoretto}, G. and {Contursi}, G. and {Cooper}, W.~J. and {Cornez}, T. and {Cowell}, S. and {Crifo}, F. and {Cropper}, M. and {Crosta}, M. and {Crowley}, C. and {Dafonte}, C. and {Dapergolas}, A. and {David}, M. and {David}, P. and {de Laverny}, P. and {De Luise}, F. and {De March}, R.},
        title = "{Gaia Data Release 3. Summary of the content and survey properties}",
      journal = {\aap},
         year = 2023,
        month = jun,
       volume = {674},
          eid = {A1},
        pages = {A1},
          doi = {10.1051/0004-6361/202243940},
archivePrefix = {arXiv},
       eprint = {2208.00211},
 primaryClass = {astro-ph.GA},
       adsurl = {https://ui.adsabs.harvard.edu/abs/2023A&A...674A...1G}
}

@Article{gladman1995,
  author = 	 {Gladman, B. J. and
	            Burns, J. A. and
							Duncan, M. J. and
							Levison, H. F.},
  title = 	 {{The Dynamical Evolution of Lunar Impact Ejecta}},
  journal = 	 {Icarus},
  year = 	 {1995},
  OPTkey = 	 {},
  volume = 	 {118},
  OPTnumber = 	 {},
  pages = 	 {302 -- 321},
  OPTmonth = 	 {},
  OPTnote = 	 {},
  OPTannote = 	 {}
  }

@Article{granvik2012,
  author = 	 {Granvik, M. and
	            Vaubaillon, J. and
							Jedicke, R.},
  title = 	 {{The population of natural Earth satellites}},
  journal = 	 {Icarus},
  year = 	 {2012},
  OPTkey = 	 {},
  volume = 	 {218},
  OPTnumber = 	 {},
  pages = 	 {262 -- 277},
  OPTmonth = 	 {},
  OPTnote = 	 {},
  OPTannote = 	 {}
  }

@Inproceedings{granvik2013,
	author = {Granvik, M. and
	          Jedicke, R. and
	          Bolin, B. and
	          Chyba, M. and
	          Patterson, G. and
	          Picot, G.},
	title  ={Earth's Temporarily-Captured Natural Satellites -- The First Step Towards Utilization of Asteroid Resources},
	year = {2013},
	booktitle = {Asteroids. Prospective Energy and Material Resources},
  editor = 	 {V. Badescu},
  pages = {151 -- 167},
  OPTorganization = {},
  publisher = {Springer},
  OPTkey = 	 {},
  OPTvolume = 	 {},
  OPTnumber = 	 {},
  OPTseries = 	 {},
  OPTtype = 	 {},
  OPTaddress = 	 {},
  OPTmonth = 	 {},
  OPTnote = 	 {},
  }

@ARTICLE{granvik2018,
       author = {{Granvik}, Mikael and {Morbidelli}, Alessandro and {Jedicke}, Robert and
         {Bolin}, Bryce and {Bottke}, William F. and {Beshore}, Edward and
         {Vokrouhlick{\'y}}, David and {Nesvorn{\'y}}, David and
         {Michel}, Patrick},
        title = "{Debiased orbit and absolute-magnitude distributions for near-Earth objects}",
      journal = {\icarus},
         year = 2018,
        month = sep,
       volume = {312},
        pages = {181-207},
          doi = {10.1016/j.icarus.2018.04.018},
archivePrefix = {arXiv},
       eprint = {1804.10265},
 primaryClass = {astro-ph.EP},
       adsurl = {https://ui.adsabs.harvard.edu/abs/2018Icar..312..181G}
}

@Article{hill1878,
  author = 	 {Hill, G. W.},
  title = 	 {{Researches in the Lunar Theory}},
  journal = 	 {American Journal of Mathematics},
  year = 	 {1878},
  OPTkey = 	 {},
  volume = 	 {I},
  OPTnumber = 	 {},
  pages = 	 {5 -- 26, 129 -- 147, 245 -- 260},
  OPTmonth = 	 {},
  OPTnote = 	 {},
  OPTannote = 	 {}
  }

@ARTICLE{ivezic2019,
       author = {{Ivezi{\'c}}, {\v{Z}}eljko and {Kahn}, Steven M. and {Tyson}, J. Anthony and {Abel}, Bob and {Acosta}, Emily and {Allsman}, Robyn and {Alonso}, David and {AlSayyad}, Yusra and {Anderson}, Scott F. and {Andrew}, John and {Angel}, James Roger P. and {Angeli}, George Z. and {Ansari}, Reza and {Antilogus}, Pierre and {Araujo}, Constanza and {Armstrong}, Robert and {Arndt}, Kirk T. and {Astier}, Pierre and {Aubourg}, {\'E}ric and {Auza}, Nicole and {Axelrod}, Tim S. and {Bard}, Deborah J. and {Barr}, Jeff D. and {Barrau}, Aurelian and {Bartlett}, James G. and {Bauer}, Amanda E. and {Bauman}, Brian J. and {Baumont}, Sylvain and {Bechtol}, Ellen and {Bechtol}, Keith and {Becker}, Andrew C. and {Becla}, Jacek and {Beldica}, Cristina and {Bellavia}, Steve and {Bianco}, Federica B. and {Biswas}, Rahul and {Blanc}, Guillaume and {Blazek}, Jonathan and {Blandford}, Roger D. and {Bloom}, Josh S. and {Bogart}, Joanne and {Bond}, Tim W. and {Booth}, Michael T. and {Borgland}, Anders W. and {Borne}, Kirk and {Bosch}, James F. and {Boutigny}, Dominique and {Brackett}, Craig A. and {Bradshaw}, Andrew and {Brandt}, William Nielsen and {Brown}, Michael E. and {Bullock}, James S. and {Burchat}, Patricia and {Burke}, David L. and {Cagnoli}, Gianpietro and {Calabrese}, Daniel and {Callahan}, Shawn and {Callen}, Alice L. and {Carlin}, Jeffrey L. and {Carlson}, Erin L. and {Chandrasekharan}, Srinivasan and {Charles-Emerson}, Glenaver and {Chesley}, Steve and {Cheu}, Elliott C. and {Chiang}, Hsin-Fang and {Chiang}, James and {Chirino}, Carol and {Chow}, Derek and {Ciardi}, David R. and {Claver}, Charles F. and {Cohen-Tanugi}, Johann and {Cockrum}, Joseph J. and {Coles}, Rebecca and {Connolly}, Andrew J. and {Cook}, Kem H. and {Cooray}, Asantha and {Covey}, Kevin R. and {Cribbs}, Chris and {Cui}, Wei and {Cutri}, Roc and {Daly}, Philip N. and {Daniel}, Scott F. and {Daruich}, Felipe and {Daubard}, Guillaume and {Daues}, Greg and {Dawson}, William and {Delgado}, Francisco and {Dellapenna}, Alfred and {de Peyster}, Robert and {de Val-Borro}, Miguel and {Digel}, Seth W. and {Doherty}, Peter and {Dubois}, Richard and {Dubois-Felsmann}, Gregory P. and {Durech}, Josef and {Economou}, Frossie and {Eifler}, Tim and {Eracleous}, Michael and {Emmons}, Benjamin L. and {Fausti Neto}, Angelo and {Ferguson}, Henry and {Figueroa}, Enrique and {Fisher-Levine}, Merlin and {Focke}, Warren and {Foss}, Michael D. and {Frank}, James and {Freemon}, Michael D. and {Gangler}, Emmanuel and {Gawiser}, Eric and {Geary}, John C. and {Gee}, Perry and {Geha}, Marla and {Gessner}, Charles J.~B. and {Gibson}, Robert R. and {Gilmore}, D. Kirk and {Glanzman}, Thomas and {Glick}, William and {Goldina}, Tatiana and {Goldstein}, Daniel A. and {Goodenow}, Iain and {Graham}, Melissa L. and {Gressler}, William J. and {Gris}, Philippe and {Guy}, Leanne P. and {Guyonnet}, Augustin and {Haller}, Gunther and {Harris}, Ron and {Hascall}, Patrick A. and {Haupt}, Justine and {Hernandez}, Fabio and {Herrmann}, Sven and {Hileman}, Edward and {Hoblitt}, Joshua and {Hodgson}, John A. and {Hogan}, Craig and {Howard}, James D. and {Huang}, Dajun and {Huffer}, Michael E. and {Ingraham}, Patrick and {Innes}, Walter R. and {Jacoby}, Suzanne H. and {Jain}, Bhuvnesh and {Jammes}, Fabrice and {Jee}, M. James and {Jenness}, Tim and {Jernigan}, Garrett and {Jevremovi{\'c}}, Darko and {Johns}, Kenneth and {Johnson}, Anthony S. and {Johnson}, Margaret W.~G. and {Jones}, R. Lynne and {Juramy-Gilles}, Claire and {Juri{\'c}}, Mario and {Kalirai}, Jason S. and {Kallivayalil}, Nitya J. and {Kalmbach}, Bryce and {Kantor}, Jeffrey P. and {Karst}, Pierre and {Kasliwal}, Mansi M. and {Kelly}, Heather and {Kessler}, Richard and {Kinnison}, Veronica and {Kirkby}, David and {Knox}, Lloyd and {Kotov}, Ivan V. and {Krabbendam}, Victor L. and {Krughoff}, K. Simon and {Kub{\'a}nek}, Petr and {Kuczewski}, John and {Kulkarni}, Shri and {Ku}, John and {Kurita}, Nadine R. and {Lage}, Craig S. and {Lambert}, Ron and {Lange}, Travis and {Langton}, J. Brian and {Le Guillou}, Laurent and {Levine}, Deborah and {Liang}, Ming and {Lim}, Kian-Tat and {Lintott}, Chris J. and {Long}, Kevin E. and {Lopez}, Margaux and {Lotz}, Paul J. and {Lupton}, Robert H. and {Lust}, Nate B. and {MacArthur}, Lauren A. and {Mahabal}, Ashish and {Mandelbaum}, Rachel and {Markiewicz}, Thomas W. and {Marsh}, Darren S. and {Marshall}, Philip J. and {Marshall}, Stuart and {May}, Morgan and {McKercher}, Robert and {McQueen}, Michelle and {Meyers}, Joshua and {Migliore}, Myriam and {Miller}, Michelle and {Mills}, David J.},
        title = "{LSST: From Science Drivers to Reference Design and Anticipated Data Products}",
      journal = {\apj},
         year = 2019,
        month = mar,
       volume = {873},
       number = {2},
          eid = {111},
        pages = {111},
          doi = {10.3847/1538-4357/ab042c},
archivePrefix = {arXiv},
       eprint = {0805.2366},
 primaryClass = {astro-ph},
       adsurl = {https://ui.adsabs.harvard.edu/abs/2019ApJ...873..111I}
}

@Article{jedicke2018,
  author = 	 {Jedicke, R. and
	            Bolin, B. T. and
							Bottke, W. F. and
							Chyba, M. and 
							Fedorets, G. and
							Granvik, M. and
							Jones, L. and
							Urrutxua, H.},
  title = 	 {{Earth's Minimoons: Opportunities for Science and Technology}},
  journal = 	 {Frontiers in Astronomy and Space Sciences},
  year = 	 {2018},
  OPTkey = 	 {},
  volume = 	 {5},
  OPTnumber = 	 {},
  pages = 	 {A13},
  OPTmonth = 	 {},
  OPTnote = 	 {},
  OPTannote = 	 {}
  }

@ARTICLE{kwiatkowski2009,
   author = {{Kwiatkowski}, T. and {Kryszczy{\'n}ska}, A. and {Poli{\'n}ska}, M. and 
	{Buckley}, D.~A.~H. and {O'Donoghue}, D. and {Charles}, P.~A. and 
	{Crause}, L. and {Crawford}, S. and {Hashimoto}, Y. and {Kniazev}, A. and 
	{Loaring}, N. and {Romero Colmenero}, E. and {Sefako}, R. and 
	{Still}, M. and {V\"{a}is\"{a}nen}, P.},
    title = "{Photometry of 2006 $RH_{120}$: an asteroid temporary captured into a geocentric orbit}",
  journal = {Astronomy \& Astrophysics},
     year = 2009,
    month = mar,
   volume = 495,
    pages = {967 -- 974}
}

@Article{micheli2012,
  author = 	 {Micheli, M. and
	            Tholen, D. J. and
							Elliott, G. T.},
  title = 	 {{Detection of radiation pressure acting on 2009 BD}},
  journal = 	 {New Astronomy},
  year = 	 {2012},
  OPTkey = 	 {},
  volume = 	 {17},
  number = 	 {4},
  pages = 	 {446 -- 452},
  OPTmonth = 	 {},
  OPTnote = 	 {},
  OPTannote = 	 {}
  }

@Article{micheli2013,
  author = 	 {Micheli, M. and
	            Tholen, D. J. and
							Elliott, G. T.},
  title = 	 {{2012 LA, an optimal astrometric target for radiation pressure detection}},
  journal = 	 {Icarus},
  year = 	 {2013},
  OPTkey = 	 {},
  volume = 	 {226},
  number = 	 {1},
  pages = 	 {251 -- 255},
  OPTmonth = 	 {},
  OPTnote = 	 {},
  OPTannote = 	 {}
  }

@Article{micheli2014,
  author = 	 {Micheli, M. and
	            Tholen, D. J. and
							Elliott, G. T.},
  title = 	 {{Radiation pressure detection and density estimate for 2011 MD}},
  journal = 	 {The Astrophysical Journal Letters},
  year = 	 {2014},
  OPTkey = 	 {},
  volume = 	 {788},
  number = 	 {1},
  pages = 	 {L1},
  OPTmonth = 	 {},
  OPTnote = 	 {},
  OPTannote = 	 {}
  }

@Article{shober2019,
  author = 	 {Shober, P. M. and
              Jansen-Sturgeon, T. and
              Sansom, E. K. and
              Devillepoix, H. A. R. and
              Bland, P. A. and
              Cup\'{a}k, M. and
              Towner, M. C. and
              Howie, R. M. and
              Hartig, B. A. D.},
  title = 	 {{Identification of a Minimoon Fireball}},
  journal = 	 {The Astronomical Journal},
  year = 	 {2019},
  OPTkey = 	 {},
  volume = 	 {158},
  OPTnumber = 	 {},
  pages = 	 {183},
  OPTmonth = 	 {},
  OPTnote = 	 {},
  OPTannote = 	 {}
  }

@INPROCEEDINGS{tody1986,
       author = {{Tody}, Doug},
        title = {{The IRAF Data Reduction and Analysis System}},
    booktitle = {Instrumentation in astronomy VI},
         year = 1986,
       editor = {{Crawford}, David L.},
       series = {Society of Photo-Optical Instrumentation Engineers (SPIE) Conference Series},
       volume = {627},
        month = jan,
        pages = {733},
          doi = {10.1117/12.968154},
       adsurl = {https://ui.adsabs.harvard.edu/abs/1986SPIE..627..733T}
}

@INPROCEEDINGS{tody1993,
       author = {{Tody}, Doug},
        title = {{IRAF in the Nineties}},
    booktitle = {Astronomical Data Analysis Software and Systems II},
         year = 1993,
       editor = {{Hanisch}, R.~J. and {Brissenden}, R.~J.~V. and {Barnes}, J.},
       series = {Astronomical Society of the Pacific Conference Series},
       volume = {52},
        month = jan,
        pages = {173},
       adsurl = {https://ui.adsabs.harvard.edu/abs/1993ASPC...52..173T}
}

@inproceedings{farnocchia2015,
       author = {{Farnocchia}, D. and {Chesley}, S.~R. and {Milani}, A. and
         {Gronchi}, G.~F. and {Chodas}, P.~W.},
        title = "{Orbits, Long-Term Predictions, Impact Monitoring}",
    booktitle = {Asteroids IV},
      editor = 	 {Michel, P. and
                  DeMeo, F. and
                  Bottke, W. F.},
         year = {2015},
        pages = {815-834},
        pulisher = {University of Arizna Press},
          doi = {10.2458/azu_uapress_9780816532131-ch041},
       adsurl = {https://ui.adsabs.harvard.edu/abs/2015aste.book..815F}
}

\begin{appendix}

\end{appendix}
\end{document}